\documentclass[final,1p,times]{elsarticle} 
\usepackage{graphicx} 
\usepackage{amssymb} 
\usepackage{amsthm} 
\usepackage{lineno} 

\journal{Nuclear Physics A} 

\newcommand{\be}{\begin{eqnarray}}
\newcommand{\ee}{\end{eqnarray}}

\begin{document} 

\begin{frontmatter} 


\title{Quarkonium Spectral Functions}

\author{\'Agnes M\'ocsy}

\address{Department of Mathematics and Science, Pratt Institute, 
Brooklyn, NY 11205, USA}

\begin{abstract} 
In this talk I summarize the progress achieved in recent years on the understanding of quarkonium properties at finite temperature. Theoretical studies from potential models, lattice QCD, and effective field theories are discussed. I also highlight a bridge from spectral functions to experiment.  
\end{abstract} 

\end{frontmatter} 


\section{Introduction}

It has been 23 years since what is today the most
cited paper in heavy ion physics \cite{matsui} was publised. In this work Matsui and
Satz conclude that "$J/\psi$ suppression in nuclear
collisions should provide an unambiguous signature of quark-gluon
plasma formation". This is how the story of quarkonium at finite
temperature began. One of the important and necessary consequences of this story is the quest for determining  properties of quarkonium states at finite temperature. Quarkonium properties can be
conveniently studied through spectral functions. 
In this talk I discuss properties of quarkonium inside an equilibrated plasma, as obtained from the
three main theoretical approaches for determining the spectral functions. These are, in
historic order, potential models, lattice QCD, and effective
field theories. I then discuss what possible implications for the
experiments the results might have.

Quarkonium states, made of a heavy quark
and its antiquark ($m_Q \gg \Lambda_{QCD}$), are tightly bound; their size is much smaller than the typical hadronic
scale, $r_{J/\psi}\simeq 0.4~$fm and $r_\Upsilon\simeq 0.2~$fm $\ll
1~$fm. Due to the heavy quark mass the quark velocity is very small,
$v\ll 1$, which allows for a non-relativistic treatment of the bound
states. Properties of quarkonium have been customarily obtained by
solving the Schr\"odinger equation, where the interaction between the
quark $Q$ and antiquark $\overline{Q}$ at distance $r$ away from each other is implemented as a potential, 
$V(r)$. This prescription works very well at zero temperature and it has
been somewhat naively used also at finite temperature. There, the idea is
that in a high temperature deconfined medium there is a
rearrangement of color around the heavy quarks, due to which the
effective charge of the $Q$ and $\overline{Q}$ is reduced. So 
light quarks and gluons screen the potential between the $Q$ and the
$\overline{Q}$, analogous to the known Debye-screening in QED plasmas. The original Matsui-Satz idea is that when the range of screeing becomes smaller than the size of
the state, $r_D<r_{Q\overline{Q}}$, the quark and antiquark cannot see each other and the two
are no longer bound. Customarily, these effects were 
described with a temperature-dependent potential $V(r,T)$.

This simple and attractive idea leaves a few questions: How good is
it to describe medium effects with a temperature dependent potential? 
Is there Debye-screening in the plasma? Is screening the
only relevant effect? And what do we mean by a bound state at finite
temperature? 

\section{Spectral Functions}

Quarkonium properties are conveniently described using
spectral functions. These contain all the relevant information about a given 
channel, providing a unified treatment of bound-states, continuum, and
threshold. Bound states show up as peaks in the spectral
function. At zero temperature states are well
defined, the widths of the peaks can be narrow, the lifetimes of the states can be large.
 The energy gap between the continuum threshold and a peak
position defines the binding energy, $E_{bin}$ for that state. 
Peaks are expected to broaden with
increasing temperature. The disappearance of a peak from the spectral
function means the bound state is dissociated (often referred to as the state is melted). Here I want to emphasize, that bound states may disappear well before the binding energy vanishes. Thus a melting (or dissociation) temperature should be defined by a less severe condition than zero binding energy, contrary to what has generally been considered in the literature.

There are three main theoretical approaches to determine spectral
functions. 1) Finite temperature potential models date back to the Matsui and Satz paper. 
Besides they are easy to handle, a great advantage of the newer versions of potential models is that they easily accomodate
non-perturbative information from lattice QCD calculations of the free energy of static quarks, in form of different "lattice-based potentials". The drawback is, however, their ad-hoc nature. 
 2) The first lattice QCD
calculations for quarkonium spectral functions at finite temperature appeared in
2003. Although the computation of quarkonium correlators is exact, it is really difficult to extract spectral functions on quantitative and sometimes even qualitative level from these. 
3) Effective field theories  have only appeared in the last few years. These are obtained
directly from QCD and they can clarify the domain of applicability for the
simpler potential model approach. However, currently derivations of the potentials from effective theories must assume some scale separation (between the bound state- and temperature-scales), and utilize
weak-coupling techniques, which makes their application near the
critical temperature problematic. 
\begin{figure*}[h]
\centering
\begin{center}
\resizebox{0.9\textwidth}{!}{%
\includegraphics{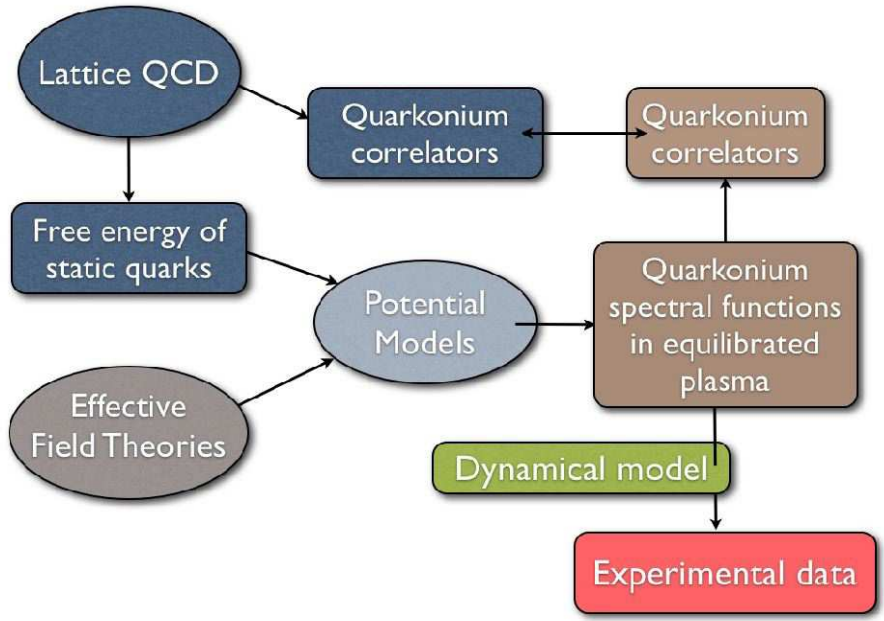}
}%
\end{center}
\caption[]{Flow chart of in-mdeium quarkonium calculations.}
\label{fig:chart}
\end{figure*} 
With the advantages and draw-backs of each of these approaches, a combination of
all their results provides a newly emerging complete picture of quarkonium at
finite temperature. The chart on figure \ref{fig:chart} illustrates the interconnectedness of these approaches.

\section{Color Screening from Lattice QCD}

\begin{figure}[h]
\begin{center}
\includegraphics[width=6.3cm]{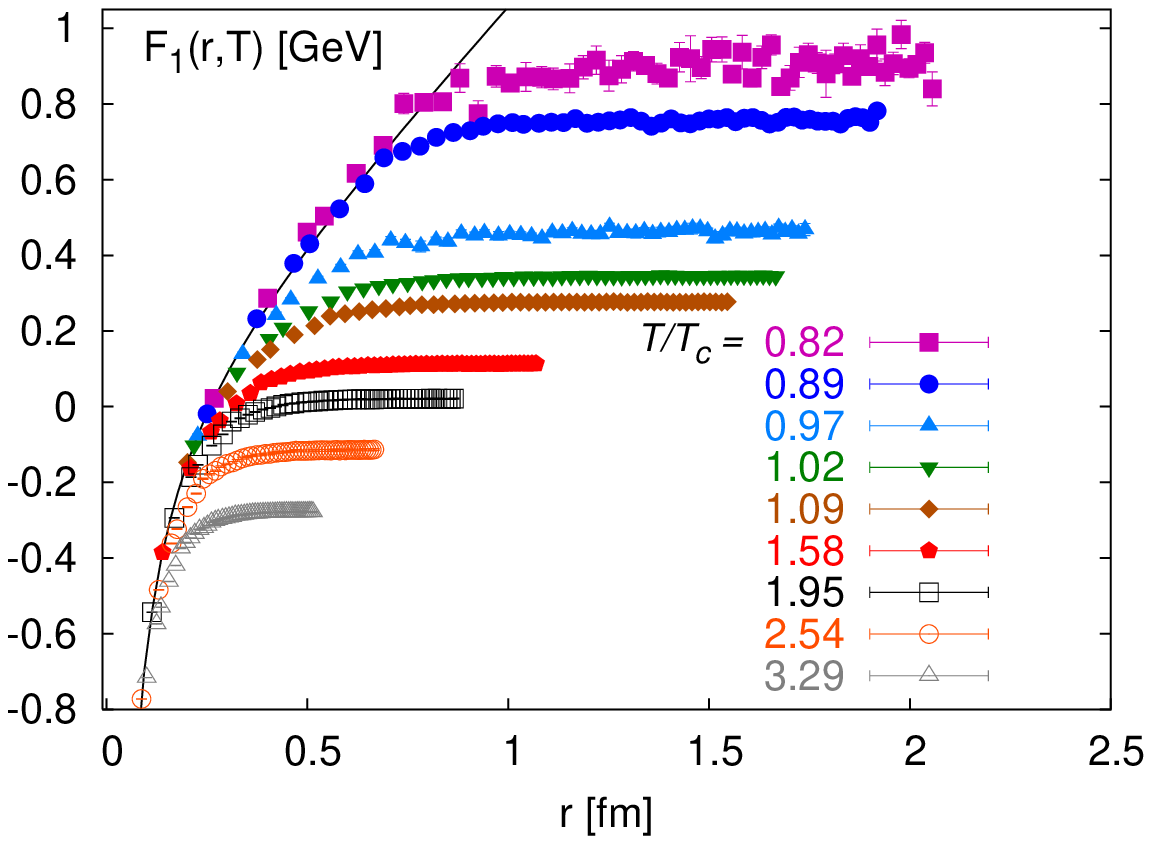}
\includegraphics[width=6.7cm]{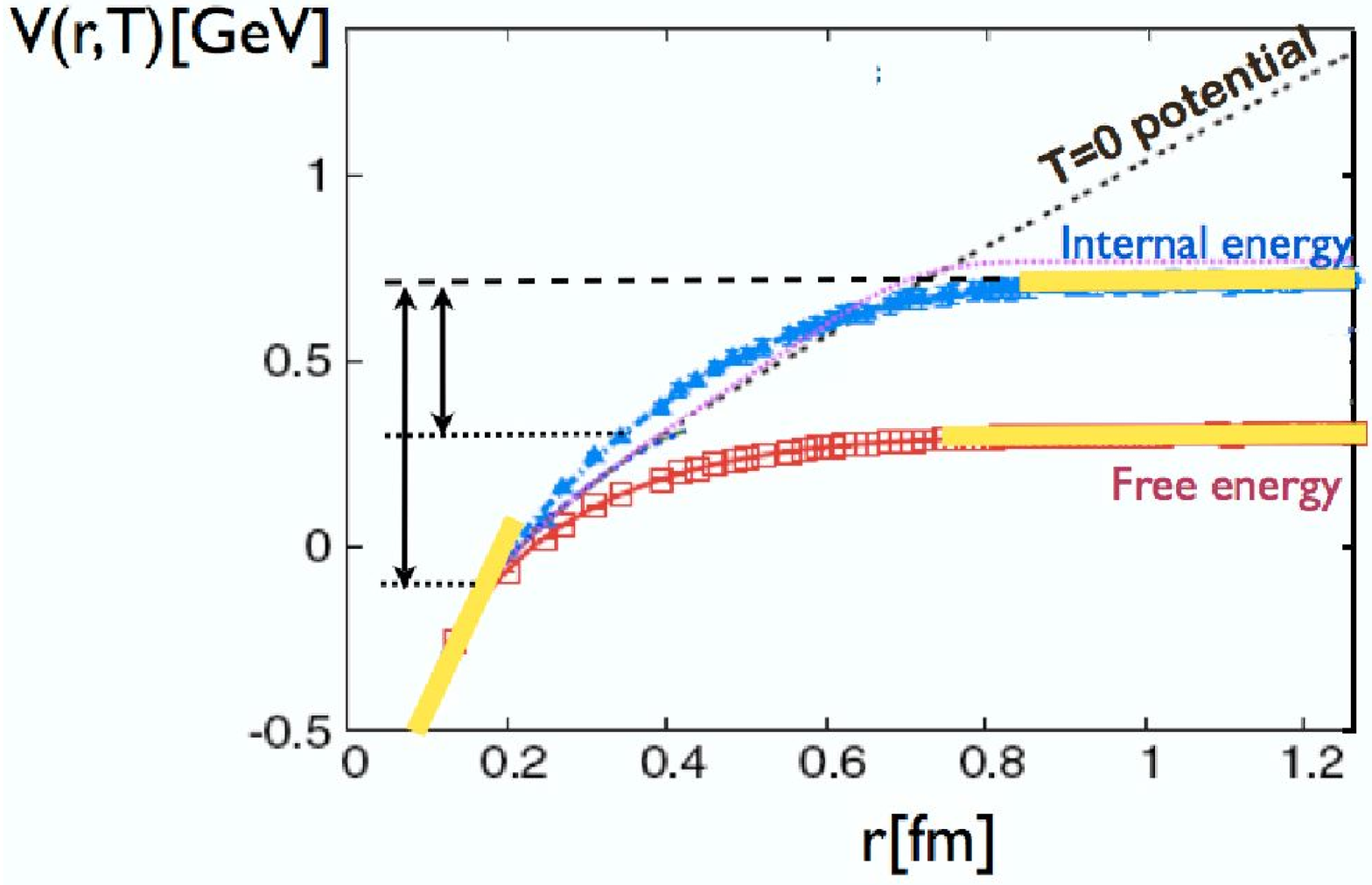}
\caption[]{Heavy quark singlet free energy versus quark separation calculated in 2+1 flavor QCD
 for different temperatures  \cite{kaczmarek}(left) and constraints from lattice on the potential at a given temperature (right).}
\label{fig:f1}
\end{center}
\end{figure}

A rigorous way to study the modification of inter-quark forces with temperature $T$ and quark separation $r$ is provided from lattice QCD by computing the heavy quark singlet free-energy, $F_1(r,T)$  \cite{kaczmarek}. This is shown in the left panel Figure \ref{fig:f1}.  At short distances, $r<r_{med}(T)$,  the singlet free
energy is independent of temperature and coincides with the zero temperature potential. The range $r_{med}(T)$ where  $F_{1}(r,T)=F_{1}(r)$ decreases with increasing temperature. At large distances $r>r_{scr}(T)$ and temperatures above the critical $T_c$ the singlet free-energy is screened, $F_{1}(r,T)=F_{1}(T)$. With increasing $T$ the strong modification of the static $Q-\overline{Q}$ interaction sets in at shoter and shorter distances $r_{scr}(T)$.
As illustrated on the right panel of Figure \ref{fig:f1} the range of $Q-\overline{Q}$ interaction is comparable to the size of $Q\overline{Q}$ charmonium states. This generic observation together with the original Matsui-Satz argument suggests that the charmonium states, such as the $J/\psi$, and excited bottomonium states, do not exist as bound states above $T_c$, while the ground state bottomonium, $\Upsilon$, is small enough to  survive deconfinement up to higher temperatures. One problem with this line of reasoning is that the singlet free energy, $F_1(r,T)$, in general, does not coincide with the potential $V(r,T)$. It has thus been argued to consider the internal energy, obtained from the free energy, as the heavy quark potential. However, the internal energy is also not the potential. The right panel of Figure \ref{fig:f1} further illustrates that at a given temperature the internal energy is more binding than the corresponding free energy, leading possibly to survival of quarkonium states when identified as the heavy quark potential. In what follows, I will discuss spectral functions obtained from a variety of lattice-based potentials, and argue that survival of the $J/\psi$ well above $T_c$ is unlikely.    

\section{Spectral Functions from Potential Models}

In potential models the basic assumption is that all medium effects are accounted for as a temperature-dependent potential. The potential that describes heavy quark-antiquark interaction in a hot medium is not yet known. Furthermore,  even the applicability domain of the traditional potential model approach is questionable. 
The newest advancements towards deriving the potential from QCD are from effective field theories, discussed in a following section.  
Lacking an exact form, potentials are either phenomenological \cite{kms} or lattice-based \cite{digal,mocsyold,wong,alberico,rapp,mocsyPRD,mocsyPRL}. 

The main advantage of lattice-based potentials is that non-perturbative effects can be conveniently accommodated using lattice data on the heavy quark-antiquark singlet free energy discussed above and shown in the left panel of Fig.~\ref{fig:f1}. 
The true potential at a given temperature and large distances (if it can be addressed at all in this approach) lies somewhere between the free- and the internal-energy (see right panel of Fig.~\ref{fig:f1}). As said in the previous section, at short distances it corresponds to the zero temperature potential, but we do not have information about its behavior at intermediate distances. Utilizing these constraints one can interpolate at intermediate distances and construct lattice-based potentials. Clearly, the shortcoming of this approach is its somewhat ad hoc nature. We can however, look at the most binding potential still consistent with lattice and learn about the quarkonium spectral functions in this upper limit. 

\begin{figure}[htbp]
\begin{center}
\includegraphics[width=7cm]{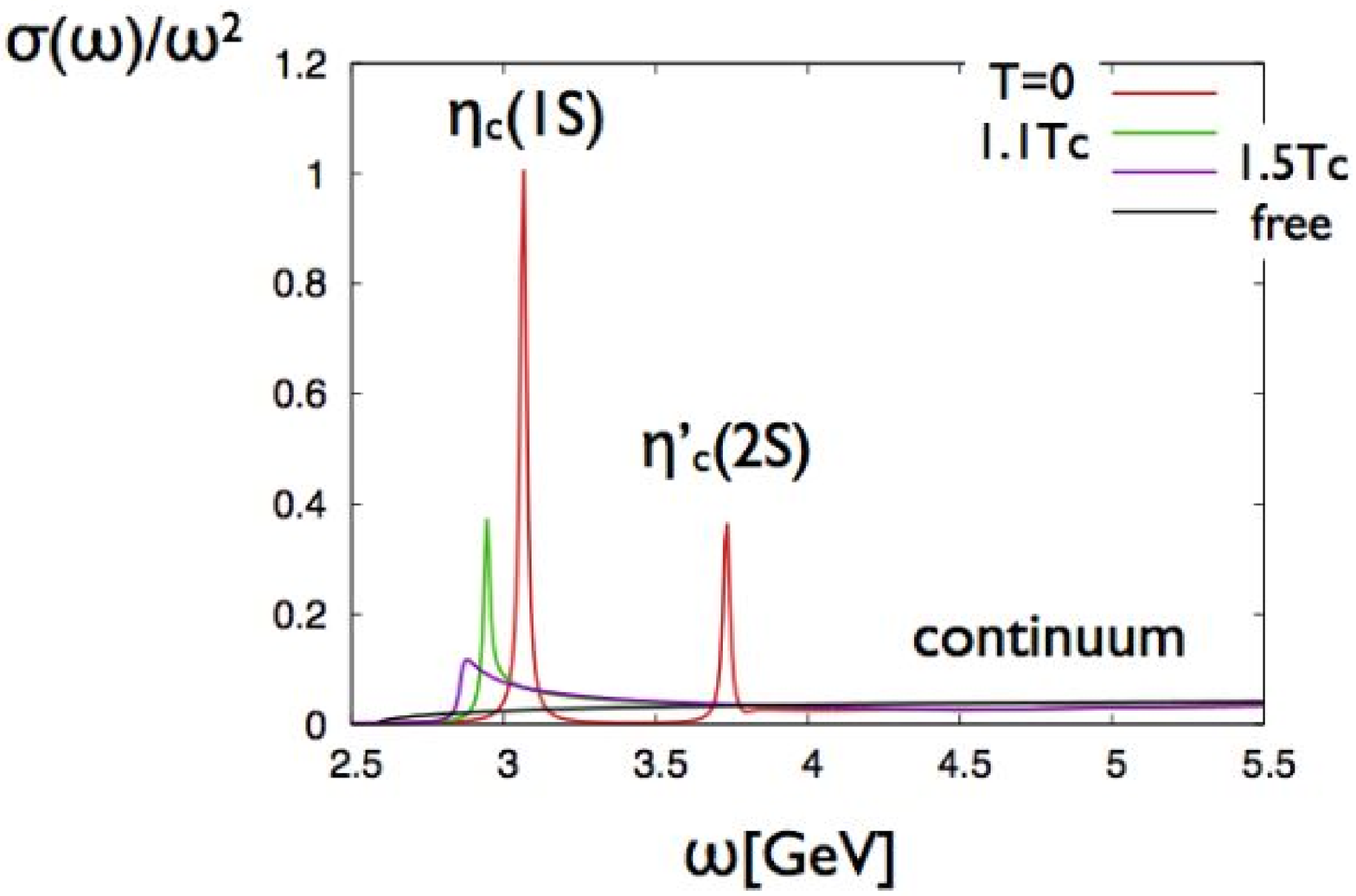}
\includegraphics[width=6cm]{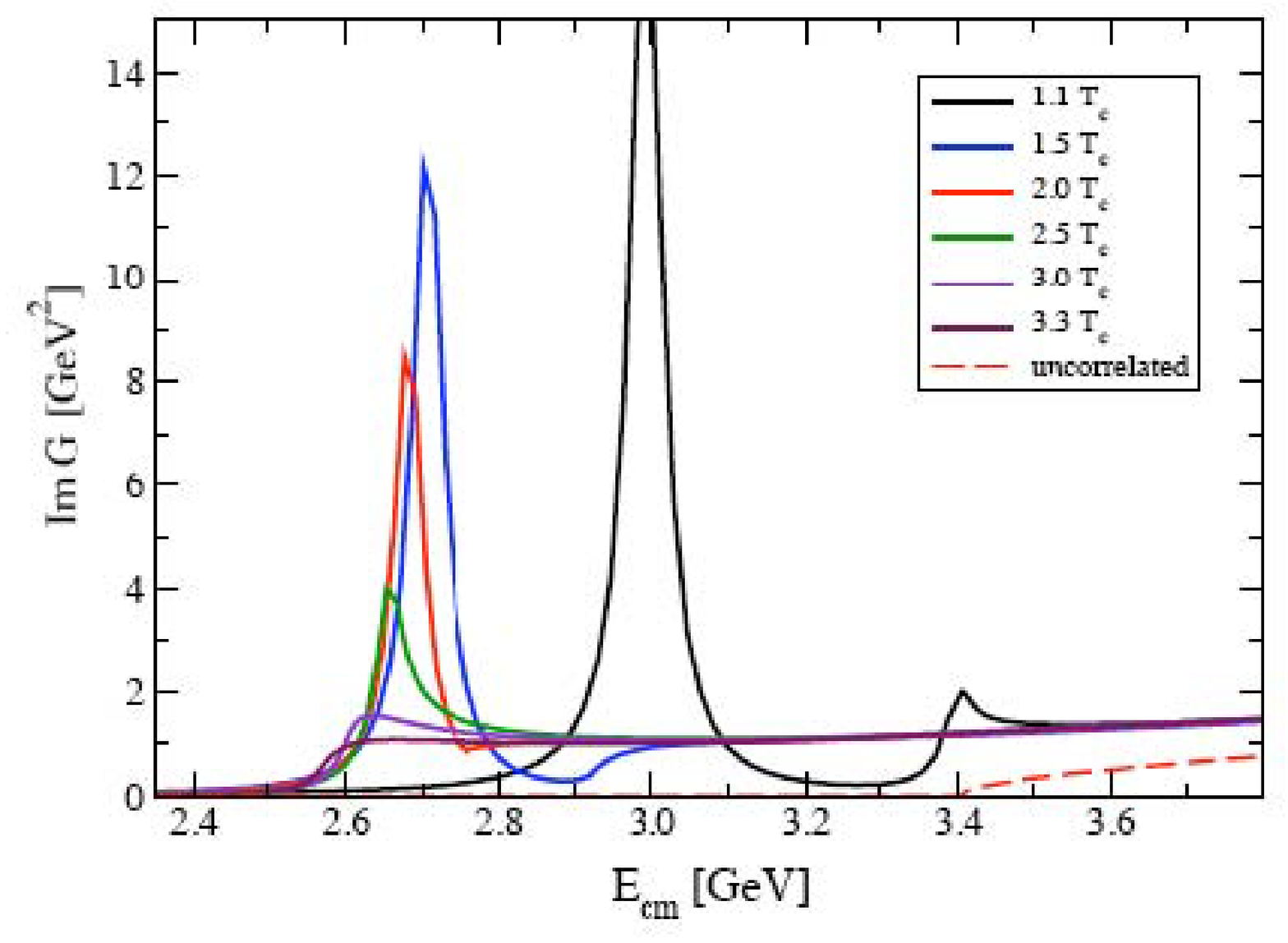}
\caption{S-wave charmonium spectral functions calculated in potential model from \cite{mocsyPRL} (left) and from \cite{rapp} (right).}
\label{fig:spf} 
\end{center}
\end{figure}

Figure \ref{fig:spf} shows quarkonium spectral functions obtained from a non-relativistic Greens function
calculation using the most confining potential in full QCD \cite{mocsyPRL} (left panel) and from a T-matrix calculation using the internal energy as potential in quenched QCD \cite{rapp} (right panel). Both results are for the S-wave charmonium channel. 
The spectral function in the left panel  shows no peak structures at high temperature, indicating that all charmonium states
are dissolved in the deconfined phase \cite{mocsyPRD,mocsyPRL}. The result shown in the right panel  finds the ground state peak  at somewhat higher temperatures \cite{rapp}. 

Even though there are some differences in the details, there are some essential features common to all models:  
For one, there is a large threshold (rescattering) enhancement beyond what corresponds to free quark propagation. This enhancement is present even at high temperatures and it 
indicates correlations persisting between the quark and antiquark even in the absence of bound states. Threshold enhancement has been identified in all of the channels (charmonium and bottomonium S- and P-states) \cite{mocsyPRD,mocsyPRL}. 
Second, there is a strong decrease with increasing temperature of the binding energies (the distance between peak position and continuum threshold in the spectral function) determined from potential models \cite{mocsy}. The decreased binding energy implies a large increase in the thermal width as well. As mentioned earlier in this talk, bound states disappear before the binding energy goes to zero \cite{mocsy}. One can think that this happens when the width becomes larger than the binding energy. In other words the time it takes for a quark and antiquark to bind is larger than the time it takes for the bound state to decay. 
\begin{figure}[htbp]
\begin{center}
\includegraphics[width=6cm]{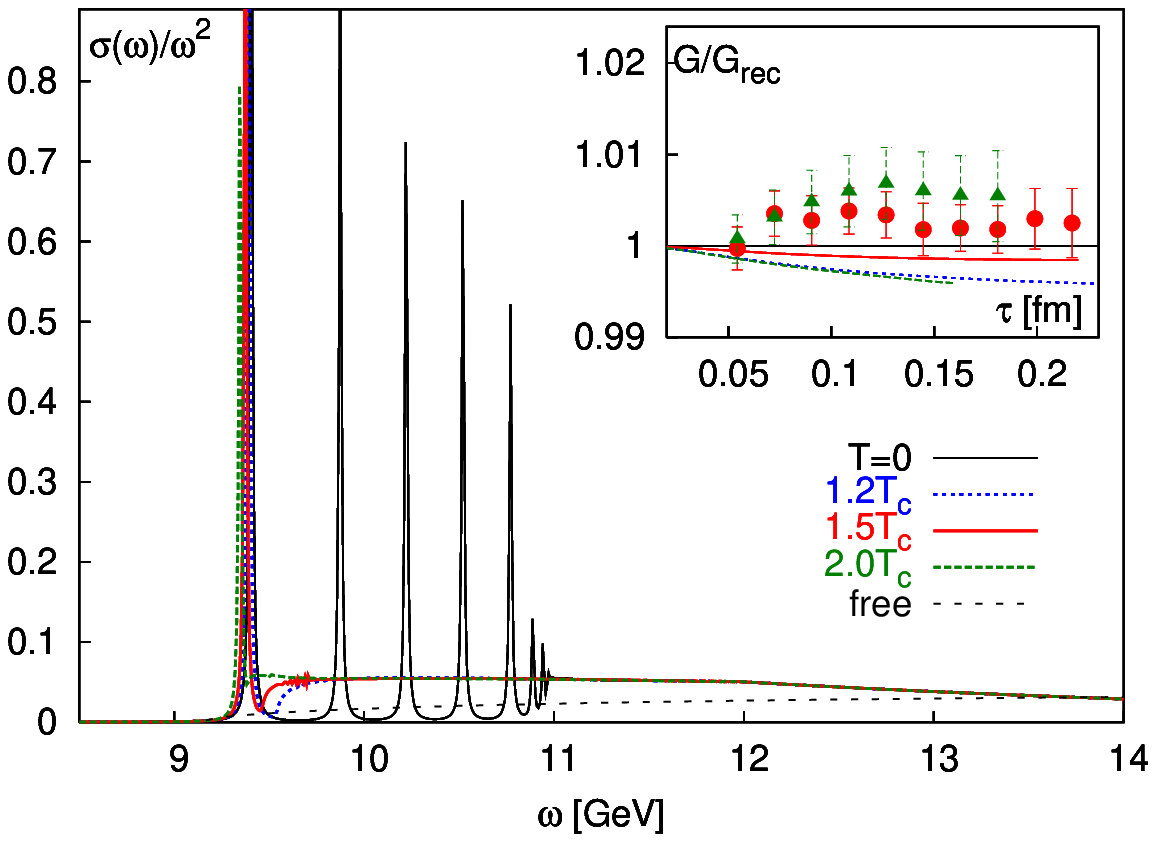}
\includegraphics[width=6cm]{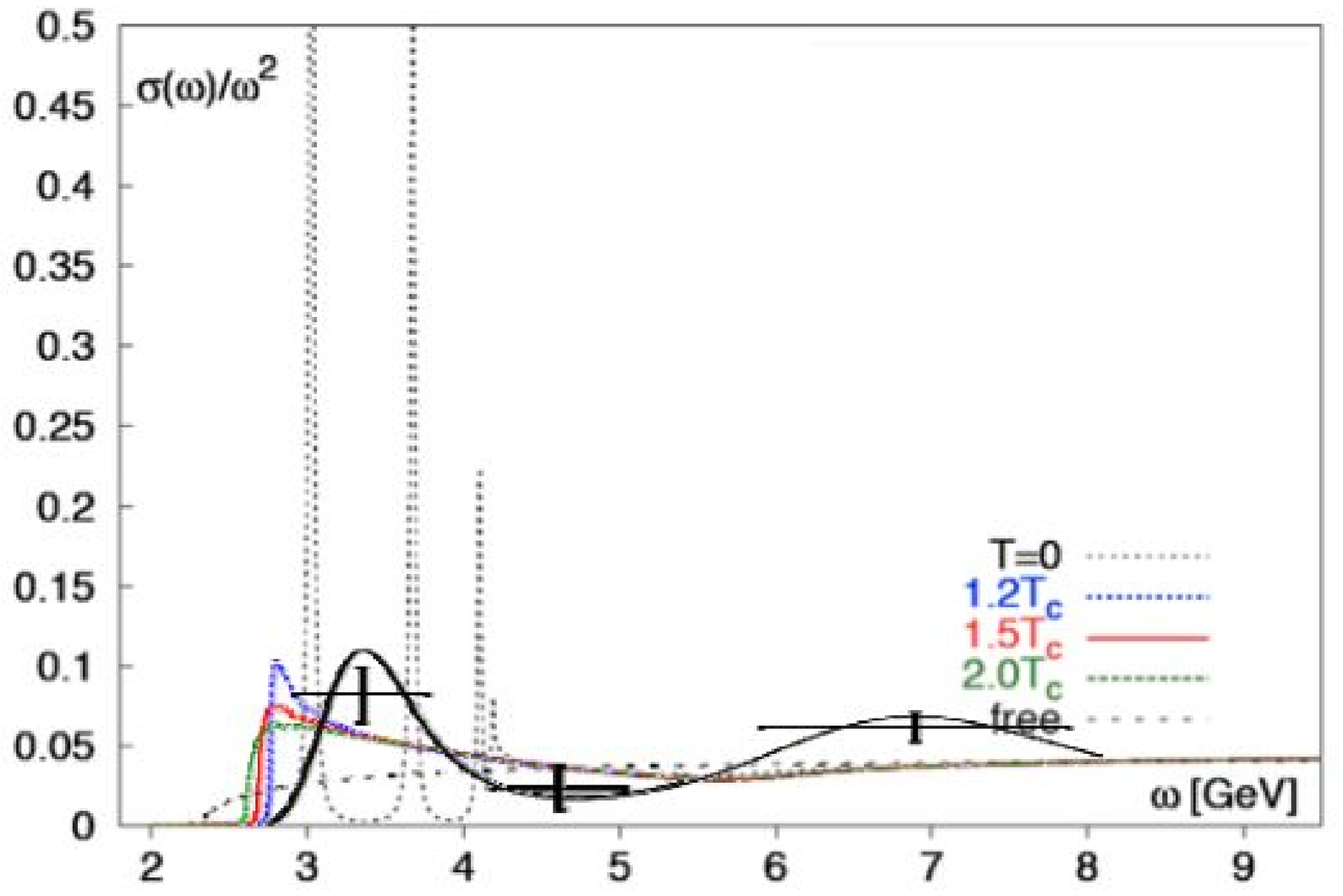}
\caption{S-wave bottomonium (left) and charmonium (right) spectral functions
calculated in potential model \cite{mocsyPRD}. Insets: correlators compared to lattice data. Charmonium is compared to lattice data from \cite{jako}}
\label{fig:spf-b} 
\end{center}
\end{figure}
Figure \ref{fig:spf-b} shows a ground state bottomonium peak up to higher temperatures into the deconfined medium, but with a largely reduced binding energy. I note that binding energies are somewhat increased in a viscous medium \cite{aniso}. For further details on anisotropy effects on quarkonium I refer the reader to \cite{dumitru}. I further note that potential model calculations of spectral functions do not include a  realistic temperature-dependent width for the states. Widths naturally arise from effective theories \cite{laine,vairo}, as discussed in a later section. We calculated the thermal widths from the binding energies \cite{mocsyPRL} following \cite{dima}. Thermal broadening of quarkonium has been addressed also in a NLO perturbative QCD \cite{park} and a QCD sum rule \cite{morita} analysis. All of these calculations show that the $J/\psi$ for example, is significantly broadened at temperatures right above that of deconfinement. 

From the analysis of the spectral functions one can provide an upper bound
for the dissociation temperatures, i.e. the temperatures above which no
bound states peaks can be seen in the spectral function. Above these temperatures bound state formation is suppressed. Conservative upper limit dissociation
temperatures for the different quarkonium states have been obtained from a full QCD calculation \cite{mocsyPRL}. Accordingly, charmonium states are dissolved in the deconfined phase, while the bottomonium ground
state may persist up to temperature of about $2T_c$. 

\section{Spectral Functions from Lattice QCD}

In lattice QCD, current-current correlation functions of mesonic currents in Euclidean-time are calculated \cite{datta,jako,aarts}. These correlators are related to the spectral function through the integral: 
\be
G(\tau,T)=\int d\omega\sigma(\omega,T)\cosh\left(\omega(\tau-1/2T)\right)/\sinh\left(\omega/2T\right)\, . \label{G}
\ee 
It is customary to present the temperature-dependence of the correlators as the ratio $G/G_{recon}$, where $G_{recon}(\tau,T=0)$ is the correlator ÒreconstructedÓ from spectral function at zero (or some low) temperature. 
Since correlators are computed with high accuracy, and 
they can be determined in potential models by integrating the spectral function using equation (\ref{G}), it has been suggested to compare correlators from potential models to the ones from lattice QCD \cite{mocsyold}. 
 Somewhat surprisingly, Euclidean correlation functions
show very little temperature dependence, irrespective of whether a state is there (such as the $\Upsilon$) or not (such as the $J/\psi$). Note also, that correlators from potential models are in accordance with the lattice calculations (see insets in Fig.~\ref{fig:spf}) (for a review see \cite{mocsy}). Originally, the small temperature
dependence of the correlators was considered as evidence for survival of different quarkonium
states \cite{datta,japan}. It is now clear that this conclusion was premature. So flat correlators do not tell about the survival or melting of states, since these kind of changes do not show up in it.  We now understand that the threshold enhancement compensates for the absence of bound states
and leads to Euclidean correlation functions which show only very small
temperature dependence \cite{mocsyPRD}. 

From the current-current correlation functions for different quarkonium channels the quarkonium spectral functions are {\it extracted} \cite{datta,japan,jako}. The extraction of spectral function using the Maximum Entropy Method is still difficult due to discretization effects, statistical errors, default model dependence (for a review see\cite{bazavov}). So the uncertainties in the spectral function are significant and details of this cannot be resolved. Moreover, a seemingly surviving ground state peak is in perfect agreement with a mere threshold enhancement obtained from potential models, as shown juxtaposed on the right panel of figure \ref{fig:spf-b}. Again, a $1~$GeV wide peak obtained with huge uncertainties from lattice QCD, and peak, whose details cannot be resolved, is most likely not a true bound state peak. Improved calculations are on the way \cite{olaf}.

\section{Spectral Functions from Effective Field Theories}

In effective field theories the quarkonium potential is derived from
the QCD Lagrangian. The basis of these theories is the existence of
scales related to the bound state and scales related to the
temperature. At T=0 for heavy quarks the existence
of different energy scales related to the heavy quark mass $m$, the 
inverse size $m v \sim 1/r$ , and the binding energy $m v^2 \sim E_{bin}$ makes it possible to construct a sequence of effective theories.  The effective theory
which emerges after integrating out the scale $m$ and $m v^2$ is pNRQCD, which delivers the potential model at $T=0$ \cite{pnrqcd}. 
Recently this approach has been extended to finite temperature, where the presence of the scales $T$, 
the Debye mass $m_D \sim g T$, and the magnetic scale $g^2 T$, makes computations more difficult. With the assumption of $m \gg T$ and weak coupling $g \ll 1$ these scales are
well separated. Depending on how these thermal scales
compare with the bound state scales, the different hierarchies allow for the 
derivation of different effective theories for quarkonium bound states
at finite temperature \cite{vairo}. 
A finding common to all of the theories is that the quarkonium
potential has not just a real part but also an imaginary part. The
temperature dependent imaginary part  determines the
thermal width, and thus the smearing out, of the quarkonium states in the spectral function. 
Several physical processes
contribute to this thermal width. For instance the scattering of
particles in the medium with gluons (Landau damping) \cite{laine}, or the thermal break-up of a
color singlet $Q-\overline{Q}$ into a color octet state and
gluons (octet transition) \cite{vairo}. Smearing of states, leading to their dissolution, even before the onset of the exponential Debye-screening of the real part of the potential (see e.g. discussion in \cite{laine}).

Another discovery from the new effective field
theory calculations is that finite temperature effects can be other
than the originally thought exponential screening of the real part of the potential; in the weak coupling approach thermal
corrections to the potential are obtained only when the
temperature is larger than the binding energy \cite{vairo}.

The current shortcomings of effective field theories are that near $T_c$ the applicability of weak coupling
techniques is problematic and theories must incorporate also non-perturbative effects.

\section{Relevance for experiments}

Knowing the quarkonium spectral functions in equilibrium QCD is necessary, but alone
is not sufficient to predict their production in heavy ion collisions. In principle, there is a simple relation between quarkonium spectral functions and quarkonium production rates measurable in experiments. This relation will hold only in thermal equilibrium, which is not likely the achieved for heavy quarks in heavy ion collisions. The bridge between theoretical spectral functions and experimental yield measurements requires additional dynamical modeling (see figure \ref{fig:chart}).  There are several such models with different underlying assumptions \cite{redlich,rapp2,seq,clint}.   

\begin{figure}[htbp]
\begin{center}
\includegraphics[width=13cm]{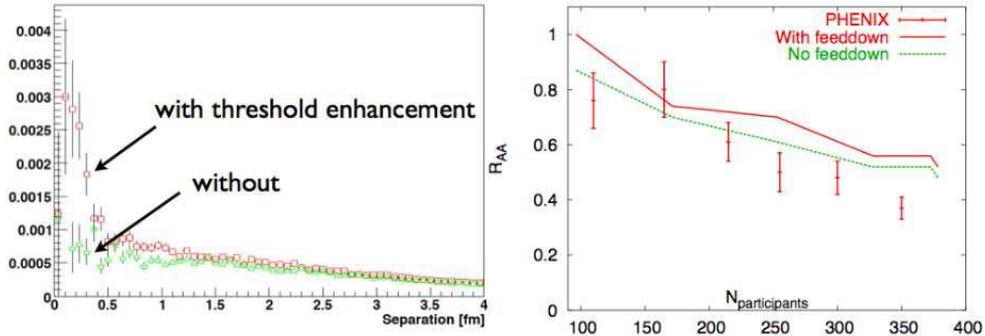}
\caption{Distribution of $c-\overline{c}$ pairs versus the inter-quark distance (left) and predicted suppression (right) from \cite{clint}.}
\label{fig:edward} 
\end{center}
\end{figure}

Let us look for instance at RHIC energies, where we learned from potential models that no $J/\psi$ peak in the spectral function is seen, only threshold enhancement \cite{mocsyPRD,mocsyPRL}. This means that 
 $J/\psi$ are not formed, only spatially correlated $c$ and $\overline{c}$ \cite{mocsyPRD,mocsyPRL}. 
 This correlation can be modeled classically, using
Langevin dynamics which includes a drag force and a random force on the heavy quark (antiquark) from the medium, as well as the forces acting between the quark and anti-quark (described by the potential). The distribution of the separation between $c-\overline{c}$ pairs with and without threshold enhancement in the spectral function is shown in the left panel of Figure \ref{fig:edward}. Some of these $c-\overline{c}$ pairs will stay correlated throughout the evolution of the system. Those pairs that are not diffused away they can bind together into a $J/\psi$ at hadronization.  
 
The right panel of Figure \ref{fig:edward} shows the result of a Langevin simulation of evolving $c-\overline{c}$ pairs on top of a hydrodynamically expanding quark-gluon plasma, which describes the RHIC data on charmonium suppression quite well \cite{clint}.  In particular, this model can explain
why, despite the fact that a deconfined medium is created at RHIC, we see
only $40-50\%$ suppression in the charmonium yield. 
%

\section{Summary}

Potential model calculations based on lattice QCD, as well as resummed perturbative QCD calculations indicate that all charmonium states and excited bottomonium states dissolve in the deconfined medium. Lattice data is consistent with $J/\psi$ screened just above $T_c$. Dissolved states lead to the reduction of the quarkonium production yield in heavy ion collisions compared to the binary-scaled proton-proton collisions. Due to possible recombination effects, however, the yield will not be zero. Potential models indicate the decrease in the binding energy of quarkonium states and a sizable threshold enhancement leading to residual quark-antiquark correlations persisting at high temperatures.  Implication of this for the understanding of experimental data can be investigated in dynamical models, which suggest that recombination of residually correlated pairs is possible when the temperature cools down sufficiently. 
Effective field theory calculations indicate temperature-dependent widths and temperature effects beyond what 
potential models have accounted for.


\section*{Acknowledgments} 

I thank the Organizers of Quark Matter 2009 for inviting me to give this talk. 
 
\end{document}